\documentclass[review,screen,anonymous,10pt]{IEEEtran}
\IEEEoverridecommandlockouts
\usepackage{cite}
\usepackage{xcolor,colortbl}

\def\BibTeX{{\rm B\kern-.05em{\sc i\kern-.025em b}\kern-.08em
		T\kern-.1667em\lower.7ex\hbox{E}\kern-.125emX}}

\usepackage[utf8]{inputenc}
\usepackage[T1]{fontenc}
\usepackage{amsmath,amssymb,amsfonts}
\usepackage[abbreviations]{foreign}
\usepackage{algorithm}
\usepackage{algpseudocode}
\usepackage{graphicx}
\usepackage[margin=3pt,skip=3pt,belowskip=-10pt,font={small,stretch=0.9},labelfont=bf]{caption}
\usepackage[subtle]{savetrees}

\usepackage[compact]{titlesec}
\usepackage{microtype}
\usepackage{setspace}
\usepackage{listings}
\usepackage{multirow}
\usepackage{scrextend}

\usepackage{subcaption}
\usepackage{textcomp}
\usepackage{booktabs}
\usepackage{xspace}
\usepackage{url}
\usepackage{pifont}
\usepackage[colorlinks=true,linkcolor=blue,anchorcolor=blue,citecolor=blue,filecolor=black,menucolor=black,runcolor=black,urlcolor=blue]{hyperref}
\usepackage{lipsum}

\usepackage[inline]{enumitem}
\setlist{noitemsep,topsep=0pt,parsep=0pt,partopsep=0pt}

\usepackage[capitalise]{cleveref}

\usepackage{tikz}

\usepackage{multicol}
\definecolor{lightcolor}{rgb}{0,0.5,1}

\usepackage{fontawesome5}
\newboolean{showcomments}
\setboolean{showcomments}{true}
\ifthenelse{\boolean{showcomments}}
{ \newcommand{\mynote}[3]{
		\fbox{\bfseries\sffamily\scriptsize#1}
		{\small$\blacktriangleright$\textsf{\emph{\color{#3}{#2}}}$\blacktriangleleft$}}}
{ \newcommand{\mynote}[3]{}}

\definecolor{darkgreen}{rgb}{0.3,0.5,0.3}
\definecolor{darkblue}{rgb}{0.3,0.3,0.5}
\definecolor{darkred}{rgb}{0.5,0.3,0.3}

\lstdefinestyle{CStyle}{
	backgroundcolor=\color{backgroundColour},   
	commentstyle=\color{mGreen},
	keywordstyle=\color{magenta},
	numberstyle=\tiny\color{mGray},
	stringstyle=\color{mPurple},
	basicstyle=\footnotesize,
	breakatwhitespace=false,         
	breaklines=true,                 
	captionpos=b,                    
	keepspaces=true,                 
	numbers=left,                    
	numbersep=5pt,                  
	showspaces=false,                
	showstringspaces=false,
	showtabs=false,                  
	tabsize=2,
	language=C
}

\newcounter{numobserv} 
\definecolor{beaublue}{rgb}{0.88, 0.93, 0.93}
\usepackage{tcolorbox}
\colorlet{shadecolor}{beaublue}
\tcbset{
	colback=beaublue,
	boxrule=0.5pt,
	boxsep=0pt,
	left=1pt,
	right=1pt,
	top=1pt,
	bottom=1pt,
	sharp corners,
	colframe=white,
}

\newcommand{\copyrightnotice}{\begin{tikzpicture}[remember picture,overlay]       
	\node[anchor=south,yshift=2pt,fill=yellow!20] at (current page.south) {\fbox{\parbox{\dimexpr\textwidth-\fboxsep-\fboxrule\relax}{\copyrighttext}}};
	\end{tikzpicture}
}
\newcommand{\copyrighttext}{  \scriptsize \textcopyright 2023 IEEE.               
	Personal use of this material is permitted.                                 
	Permission from IEEE must be obtained for all other uses,                   
	in any current or future media, including reprinting/republishing this      
	material for advertising or promotional purposes, creating new collective   
	works, for resale or redistribution to servers or                           
	lists, or reuse of any copyrighted component of this work in other works.   
	Pre-print version. Published in the 53rd Annual IEEE/IFIP International Conference on Dependable Systems and Networks (DSN) Doctoral Forum.
	
}

\begin{document}
\title{Enhancing IoT Security and Privacy with Trusted Execution Environments and Machine Learning}

\author{\IEEEauthorblockN{Peterson Yuhala\\}
	\IEEEauthorblockA{\textit{University of Neuchâtel, Switzerland} \\	
		peterson.yuhala@unine.ch}}

\maketitle
\copyrightnotice
	
\begin{abstract} 

With the increasing popularity of Internet of Things (IoT) devices, security concerns have become a major challenge: confidential information is constantly being transmitted (sometimes inadvertently) from user devices to untrusted cloud services. 
This work proposes a design to enhance security and privacy in IoT based systems by isolating hardware peripheral drivers in a trusted execution environment (TEE), and leveraging secure machine learning classification techniques to filter out sensitive data, \eg speech, images, \etc from the associated peripheral devices before it makes its way to an untrusted party in the cloud.
\end{abstract}
\section{Problem definition}
There has been a rapid growth of Internet of Things (IoT) platforms in recent years, with a significant part involving smart home based systems, \eg Amazon Alexa~\cite{lopatovska2019talk}, Google Home~\cite{googlehome-vr}, \etc. While these platforms allow for greater convenience and energy efficiency, their increasing popularity raises serious security and privacy concerns~\cite{apthorpe2017smart}. Large quantities of potentially security sensitive data are constantly transferred from hardware peripheral devices (\eg cameras, microphones) on the user end to untrusted cloud service providers, oftentimes with little regard to the confidentiality of the shared data. On the one hand, sensitive user information, \eg images, speech, \etc, could be involuntarily leaked to an untrusted party in the cloud, \ie Amazon, Google. For example, in July 2019, more than 1000 Google Assistant recordings were involuntarily leaked~\cite{google-leak,cnbc}, with part of these recordings activated accidentally by users~\cite{bbc-google-leak}.
On the other hand, privileged software like the operating system (OS) can be compromised, and thus constitutes a serious threat to data security and privacy~\cite{keystone,VanNostrand2019ConfidentialDL}.

Trusted execution environments (TEEs) like Arm TrustZone~\cite{pinto19} provide a hardware-based secure sandbox to shield sensitive programs. Various IoT platforms ship with Arm TrustZone enabled hardware, and the latter has been leveraged by research and industry to secure sensitive programs~\cite{gottel19}. However, there is still a lack of concrete solutions addressing the security and privacy of data from hardware peripherals which constitute an IoT setup like a smart home. In this work, we leverage TEE technology and machine learning (ML) techniques to prevent involuntary exposure of sensitive peripheral data to an untrusted cloud provider or OS.

\section{Proposed solution}
Our design constitutes securing hardware peripheral device driver software with Arm TrustZone, and leveraging ML classification techniques to filter out sensitive information being sent to an untrusted party like Amazon or Google. Our design is based on OP-TEE, an open source TEE implementation for securing applications based on TrustZone technology. OP-TEE is designed as a companion software to a non-secure Linux kernel running on ARM based devices~\cite{optee}, and secures \textit{trusted applications} (TAs) from the non-secure OS, as well as other TAs. OP-TEE provides a secure interface called a  \textit{pseudo trusted application} (PTA)~\cite{pta} which is a secure module with OS-level privileges that could serve as an intermediary between a TA (no OS-level privileges) and low-level code like device driver software. 

\begin{figure}
	\centering
	\hspace*{-0.5cm} 	
	\includegraphics[scale=0.55]{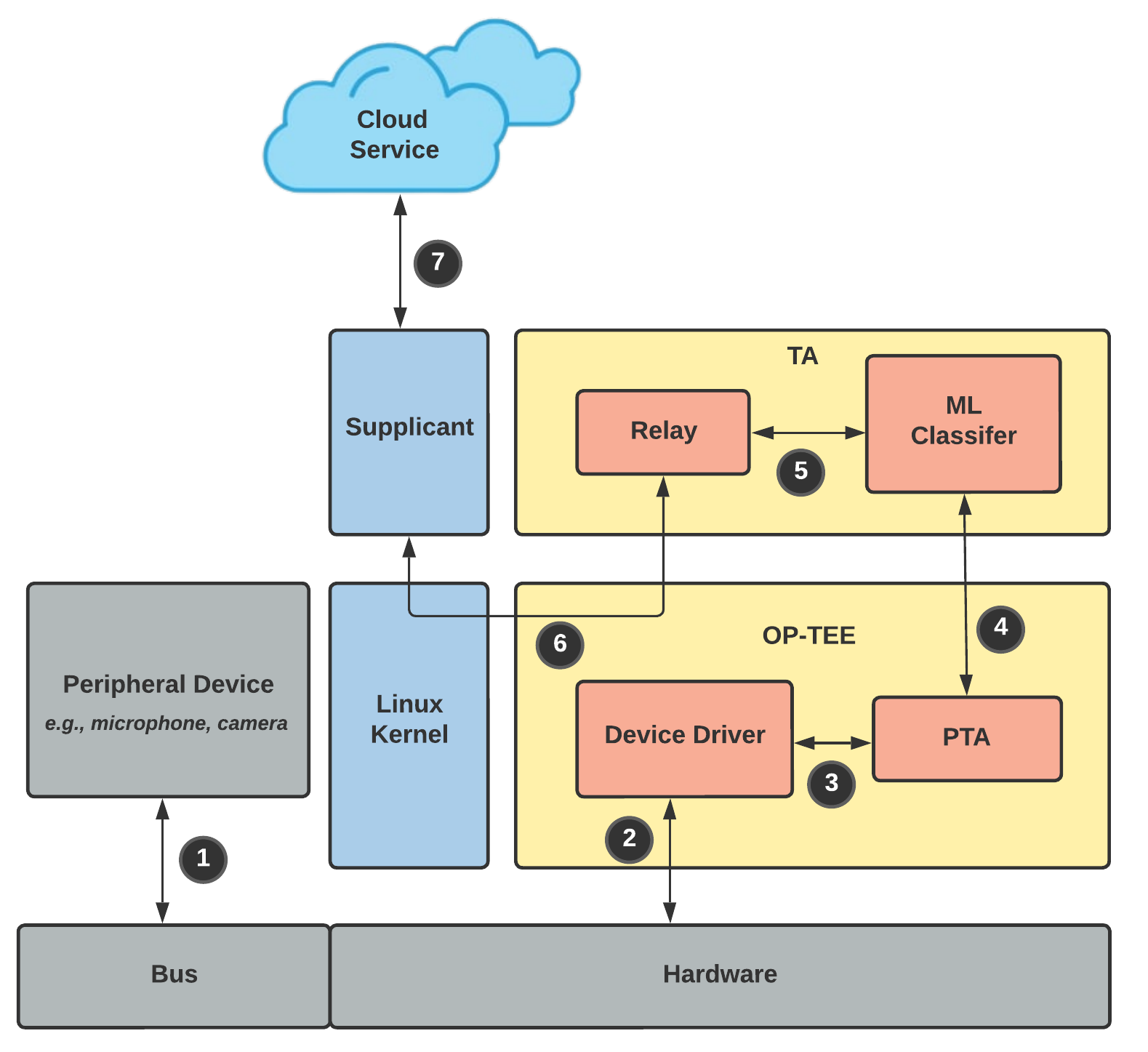}
	\caption{Securing sensitive peripheral device data with OP-TEE and ML.}
	\label{fig:architecture}
\end{figure} 
 
 The proposed design is outlined in \autoref{fig:architecture}: a peripheral device (\eg microphone, camera) which constitutes a smart home/IoT setup receives input information (\ie speech, images)\ding{202} from a user; this data could be sensitive or not. In a regular setup, the device driver software is part of the untrusted OS, thus leaking sensitive data. Our design ports the full driver software into OP-TEE. As such, the secure hardware device driver associated with the peripheral device reads this potentially sensitive data into its I/O buffers \ding{203}. TrustZone provides an address space controller capable of carving out secure RAM memory from which a secure driver's I/O buffers are allocated. The sensitive data is thus securely processed (\eg encoding an audio signal) by the driver software, after which it is transferred to a trusted application via the PTA interface \ding{204}-\ding{205}. The TA also executes in secure memory, and comprises a pre-trained ML classifier capable of determining potentially sensitive information; the latter is filtered out of the data stream \ding{206} before it is sent to an untrusted cloud service like Amazon or Google via a relay module in the TA \ding{207}-\ding{208}. The relay module leverages an OP-TEE user space daemon called the \textit{TEE supplicant} to provide OS-level services such as network communication. 
\section{Results}
So far we have begun a proof-of-concept (POC) implementation of our approach on the NVIDIA Jetson AGX Xavier development kit~\cite{jetson-xavier}. This kit ships with a TrustZone-enabled ARMv8.2 (64-bit) CPU, and integrates deep learning capabilities as well as rich I/O. Our POC focuses on inter-IC sound (I$^2$S)~\cite{philips-i2s} capable peripheral devices, like microphones~\cite{sph0645-mic}. We chose the I$^2$S protocol for our preliminary use case because it is lightweight, contrary to more complex protocols like USB.

We are yet to perform concrete experiments to evaluate our proposed design. However, we expect there will be trade-offs between security and performance. That is, embedding driver software and deep learning capabilities in a TEE on a low-power IoT device certainly leads to improved security, but this is likely to come at a cost of decreased performance, and increased power consumption.

\section{Research plan}
Our key goal is to obtain a fully working POC to validate our approach. The full research plan including past and ongoing work is as follows:
\begin{enumerate}
	\item We conducted a comprehensive review of the existing literature on IoT and cloud security with TEEs. Our primary goal was to gain a better understanding of the current state of research so as to identify the research gaps. We reviewed popular TEE technologies like Intel SGX~\cite{vcostan}, which offer strong security guarantees for a wide range of programs but are adapted for server-end applications. Some research works~\cite{brasser16, trustui, secloak} propose techniques to secure peripherals by leveraging TrustZone, however their target use-cases make them insufficient for addressing the smart home IoT security issue we identified.
	
	\item Minimizing the trusted computing base (TCB) is of primary importance in the context of TEE development. Platforms like the NVIDIA Jetson AGX Xavier provide a large set of I/O devices and driver software, sometimes for the same purpose (e.g sound recording). As a result, just part of a large driver code base could be used by a target protocol, \eg I$^2$S, and thus the full driver code need not be secured within the TEE. To reduce the TCB, we have implemented a tracing mechanism within the kernel which permits to identify a minimal set of driver functionality to be ported to OP-TEE. This tracing mechanism involves logging of driver function calls when a particular task, \eg recording a sound, is being executed. The logs are then analyzed to identify a minimal set of executed functions necessary for the task to complete. We leverage conditional compiler directives to selectively exclude driver functions which are not required for the task, from being compiled and included in the final OP-TEE image.

	\item We have begun porting I$^2$S driver software for the NVIDIA Jetson AGX Xavier platform, and have a functioning PTA interface setup to communicate with userland TAs. Once the driver is ported completely, the PTA interface will be extended to complete the communication path between the TAs and driver software.
	
	\item \emph{Machine learning model}. The architecture of the ML model to be included on the TA side depends on the type of data (\ie speech, images, \etc) being analyzed. In our preliminary implementation, we focus on I$^2$S based audio analysis. As a result, a pre-trained speech recognition model can be used to transcribe the audio signals received from the device driver; several pre-trained models exist~\cite{radford2022robust, wang2020fairseq} that can be reused for this purpose. The resulting text is then fed to the ML classifier. On the other hand, for an image analysis based system, a pre-trained ML classifier alone will be sufficient.
	
	Different machine learning architectures can be applied at the level of the ML classifier:
	\begin{itemize}
	\item \emph{Convolutional neural networks (CNNs)~\cite{abdel2014convolutional}: } This is a common approach in text classification and involves feeding the input data into a convolutional layer that learns the relevant features of the data. The output of this layer is then fed into a fully connected layer that performs a binary classification, \ie sensitive or not.
	
	\item \emph{Transformers~\cite{wolf2020transformers}:} Transformer-based ML models have shown great success in natural language processing tasks such as language translation and text classification. In the context of our research, Transformers can be used to encode the initial input data so as to learn relevant features of the data via a self-attention mechanism. The encoded representation can then be fed into a binary classification layer to identify sensitive content.
	
	\item \emph{Hybrid CNN-Transformer model:} A hybrid model that combines the strengths of CNNs and Transformers can also be used to identify sensitive information. One way to combine both is to use the CNN model as a feature extractor and the transformer as a classifier. 
	\end{itemize}
	
	Overall, the choice between these architectures will depend on the nature of input (\ie speech, images, \etc) as well as the final evaluation results obtained.

	
	\item We are conducting a preliminary analysis of the design of the relay module: this module constitutes a TLS endpoint which implements an API, \eg Amazon Alexa voice service (AVS)~\cite{alexa-voice-service} used to communicate with the cloud service provider. 
	
	\item Our final goal is to harmonize our approach so it could be applied to a larger and more generic set of peripherals and data, once the full POC I$^2$S implementation is in full operation.
\end{enumerate}
\vspace{-1.5mm}
\section{Limitations}
The main limitation of our approach is performance. On the one hand, IoT devices generally have limited processing power, making it challenging to run complex ML models. Further, TEE technologies like TrustZone provide relatively small memory resources for applications. On the other hand, securing programs within a TEE usually introduces additional overhead, \eg through contexts switches between the trusted and untrusted worlds. We aim to mitigate these issues by minimizing the TCB: minimal set of driver functions, and smaller ML models. 
\section*{Acknowledgments}
This work is supported by the VEDLIoT (Very Efficient Deep Learning in IoT) European project.


\bibliographystyle{plain}
\bibliography{min}

\end{document}